
\def\service{T}
\catcode`\@=11
\def\unredoffs{\voffset=11mm \hoffset=0.5mm}

%
\newbox\leftpage \newdimen\fullhsize \newdimen\hstitle \newdimen\hsbody
\newdimen\hdim
\tolerance=400\pretolerance=800
%
%
\newif\ifsmall \smallfalse
\newif\ifdraft \draftfalse
\newif\iffrench \frenchfalse
\newif\ifeqnumerosimple \eqnumerosimplefalse
\nopagenumbers
\headline={\ifnum\pageno=1\hfill\else\hfil{\headrm\folio}\hfil\fi}
\def\draftstart{
\magnification=1200 \unredoffs\hsize=130mm\vsize=190mm
\hsbody=\hsize \hstitle=\hsize 
\nolabels
\iffrench
\dicof
\else
\dicoa
\fi
}

\font\elevrm=cmr9

\newdimen\chapskip
\font\twbf=cmssbx10 scaled 1200
\font\ssbx=cmssbx10

\font\caprm=cmr9
\font\capit=cmti9
\font\capbf=cmbx9
\font\capsl=cmsl9
\font\capmi=cmmi9
\font\capex=cmex9
\font\capsy=cmsy9
\chapskip=17.5mm
\def\makeheadline{\vbox to 0pt{\vskip-22.5pt
\line{\vbox to8.5pt{}\the\headline}\vss}\nointerlineskip}
\font\tbfi=cmmib10
\font\tenbi=cmmib7
\font\fivebi=cmmib5
\textfont4=\tbfi
\scriptfont4=\tenbi
\scriptscriptfont4=\fivebi
\font\headrm=cmr10

\font\eightrm=cmr6
\font\sixrm=cmr5
\font\eightmi=cmmi6
\font\sixmi=cmmi5
\font\eightsy=cmsy6
\font\sixsy=cmsy5
\font\eightbf=cmbx6
\font\sixbf=cmbx5
\skewchar\capmi='177 \skewchar\eightmi='177 \skewchar\sixmi='177
\skewchar\capsy='60 \skewchar\eightsy='60 \skewchar\sixsy='60

\def\elevenpoint{
\textfont0=\caprm \scriptfont0=\eightrm \scriptscriptfont0=\sixrm
\def\rm{\fam0\caprm}
\textfont1=\capmi \scriptfont1=\eightmi \scriptscriptfont1=\sixmi
\textfont2=\capsy \scriptfont2=\eightsy \scriptscriptfont2=\sixsy
\textfont3=\capex \scriptfont3=\capex \scriptscriptfont3=\capex
\textfont\itfam=\capit \def\it{\fam\itfam\capit} 
\textfont\slfam=\capsl  \def\sl{\fam\slfam\capsl} 
\textfont\bffam=\capbf \scriptfont\bffam=\eightbf
\scriptscriptfont\bffam=\sixbf
\def\bf{\fam\bffam\capbf} 
\textfont4=\tbfi \scriptfont4=\tenbi \scriptscriptfont4=\tenbi
\normalbaselineskip=13pt
\setbox\strutbox=\hbox{\vrule height9.5pt depth3.9pt width0pt}
\let\big=\elevenbig \normalbaselines \rm}

\catcode`\@=11

\font\tenmsa=msam10
\font\sevenmsa=msam7
\font\fivemsa=msam5
\font\tenmsb=msbm10
\font\sevenmsb=msbm7
\font\fivemsb=msbm5
\newfam\msafam
\newfam\msbfam
\textfont\msafam=\tenmsa  \scriptfont\msafam=\sevenmsa
  \scriptscriptfont\msafam=\fivemsa
\textfont\msbfam=\tenmsb  \scriptfont\msbfam=\sevenmsb
  \scriptscriptfont\msbfam=\fivemsb

\def\hexnumber@#1{\ifcase#1 0\or1\or2\or3\or4\or5\or6\or7\or8\or9\or
	A\or B\or C\or D\or E\or F\fi }

\font\teneuf=eufm10
\font\seveneuf=eufm7
\font\fiveeuf=eufm5
\newfam\euffam
\textfont\euffam=\teneuf
\scriptfont\euffam=\seveneuf
\scriptscriptfont\euffam=\fiveeuf
\def\frak{\ifmmode\let\next\frak@\else
 \def\next{\Err@{Use \string\frak\space only in math mode}}\fi\next}
\def\goth{\ifmmode\let\next\frak@\else
 \def\next{\Err@{Use \string\goth\space only in math mode}}\fi\next}
\def\frak@#1{{\frak@@{#1}}}
\def\frak@@#1{\fam\euffam#1}

\edef\msa@{\hexnumber@\msafam}
\edef\msb@{\hexnumber@\msbfam}

\def\Bbb{\ifmmode\let\next\Bbb@\else
 \def\next{\errmessage{Use \string\Bbb\space only in math mode}}\fi\next}
\def\Bbb@#1{{\Bbb@@{#1}}}
\def\Bbb@@#1{\fam\msbfam#1}

\catcode`\@=12
\def\sla#1{\mkern-1.5mu\raise0.4pt\hbox{$\not$}\mkern1.2mu #1\mkern 0.7mu}
\def\Dbar{\mkern-1.5mu\raise0.4pt\hbox{$\not$}\mkern-.1mu {\rm D}\mkern.1mu}
\def\Abar{\mkern1.mu\raise0.4pt\hbox{$\not$}\mkern-1.3mu A\mkern.1mu}
\def\dicof{
\gdef\Resume{RESUME}
\gdef\Toc{Table des mati\`eres}
\gdef\soumisa{Soumis \`a:}
}
\def\dicoa{
\gdef\Resume{ABSTRACT}
\gdef\Toc{Table of Contents}
\gdef\soumisa{Submitted to}
}

\def\uniset{\rlap{\elevrm 1}\kern.15em 1}
\def\bkR{{\rm I\kern-.17em R}}
\def\bkC{{\rm \kern.24em
            \vrule width.05em height1.4ex depth-.05ex
            \kern-.26em C}}

\def\frac#1#2{{\textstyle{#1\over#2}}}

\def\leaderfill{\leaders\hbox to 1em{\hss.\hss}\hfill}
\def\saclay{\if S\service \spec \else \spht \fi}
\def\spht{
\centerline{Service de Physique Th\'eorique, CEA-Saclay}
\centerline{F-91191 Gif-sur-Yvette Cedex, FRANCE}}
\def\spec{
\centerline{Service de Physique de l'Etat Condens\'e, CEA-Saclay}
\centerline{F-91191 Gif-sur-Yvette Cedex, FRANCE}}
\def\logo{
\if S\service 
\font\sstw=cmss10 scaled 1200
\font\ssx=cmss8
\vtop{\hsize 9cm
{\sstw {\twbf P}hysique de l'{\twbf E}tat {\twbf C}ondens\'e \par}
\ssx SPEC -- DRECAM -- DSM\par
\vskip 0.5mm
\sstw CEA -- Saclay \par
}
\else 
\vtop{\hsize 9cm
}
\fi }
\catcode`\@=11
\def\deqalignno#1{\displ@y\tabskip\centering \halign to
\displaywidth{\hfil$\displaystyle{##}$\tabskip0pt&$\displaystyle{{}##}$
\hfil\tabskip0pt &\quad
\hfil$\displaystyle{##}$\tabskip0pt&$\displaystyle{{}##}$
\hfil\tabskip\centering& \llap{$##$}\tabskip0pt \crcr #1 \crcr}}
\def\deqalign#1{\null\,\vcenter{\openup\jot\m@th\ialign{
\strut\hfil$\displaystyle{##}$&$\displaystyle{{}##}$\hfil
&&\quad\strut\hfil$\displaystyle{##}$&$\displaystyle{{}##}$
\hfil\crcr#1\crcr}}\,}
\openin 1=\jobname.sym
\ifeof 1\closein1\message{<< (\jobname.sym DOES NOT EXIST) >>}\else%
\input\jobname.sym\closein 1\fi
\newcount\nosection
\newcount\nosubsection
\newcount\neqno
\newcount\notenumber
\newcount\figno
\newcount\tabno
\def\content{\jobname.toc}
\def\symbols{\jobname.sym}
\newwrite\toc
\newwrite\sym
\def\authorname#1{\maketitle{\bf #1}\smallskip}
\def\address#1{ #1\medskip}
\newdimen\hulp
\def\maketitle#1{
\edef\oneliner##1{\centerline{##1}}
\edef\twoliner##1{\vbox{\parindent=0pt\leftskip=0pt plus 1fill\rightskip=0pt
plus 1fill
                     \parfillskip=0pt\relax##1}}
\setbox0=\vbox{#1}\hulp=0.5\hsize
                 \ifdim\wd0<\hulp\oneliner{#1}\else
                 \twoliner{#1}\fi}

\def\title#1{\gdef\titlename{#1}
\maketitle{
\twbf
{\titlename}}
\vskip3truemm\vfill
\nosection=0
\neqno=0
\notenumber=0
\figno=1
\tabno=1
\def\prefix{}
\def\eqprefix{}
\mark{\the\nosection}
\message{#1}
\immediate\openout\sym=\symbols
}
\def\preprint#1{\vglue-10mm
\line{ \logo \hfill {#1} }\vglue 20mm\vfill}
\def\abstract{\vfill\centerline{\Resume} \smallskip \begingroup\narrower
\elevenpoint\baselineskip10pt}
\def\endabstract{\par\endgroup \bigskip}
\def\mktoc{\centerline{\bf \Toc} \medskip\caprm
\parindent=2em
\openin 1=\jobname.toc
\ifeof 1\closein1\message{<< (\jobname.toc DOES NOT EXIST. TeX again)>>}%
\else\input\jobname.toc\closein 1\fi
 \bigskip}
\def\section#1\par{\vskip0pt plus.1\vsize\penalty-100\vskip0pt plus-.1
\vsize\bigskip\vskip\parskip
\message{ #1}
\ifnum\nosection=0\immediate\openout\toc=\content%
\edef\ecrire{\write\toc{\par\noindent{\ssbx\ \titlename}
\string\leaderfill{\noexpand\number\pageno}}}\ecrire\fi
\advance\nosection by 1\nosubsection=0
\ifeqnumerosimple
\else \xdef\eqprefix{\prefix\the\nosection.}\neqno=0\fi
\vbox{\noindent\bf\prefix\the\nosection\ #1}
\mark{\the\nosection}\bigskip\noindent
\xdef\ecrire{\write\toc{\string\par\string\item{\prefix\the\nosection}
#1
\string\leaderfill {\noexpand\number\pageno}}}\ecrire}

\def\appendix#1#2\par{\bigbreak\nosection=0
\notenumber=0
\neqno=0
\def\prefix{A}
\mark{\the\nosection}
\message{\appendixname}
\leftline{\ssbx APPENDIX}
\leftline{\ssbx\uppercase\expandafter{#1}}
\leftline{\ssbx\uppercase\expandafter{#2}}
\bigskip\noindent\nonfrenchspacing
\edef\ecrire{\write\toc{\par\noindent{{\ssbx A}\
{\ssbx#1\ #2}}\string\leaderfill{\noexpand\number\pageno}}}\ecrire}%

\def\subsection#1\par {\vskip0pt plus.05\vsize\penalty-100\vskip0pt
plus-.05\vsize\bigskip\vskip\parskip\advance\nosubsection by 1
\vbox{\noindent\it\prefix\the\nosection.\the\nosubsection\
\it #1}\smallskip\noindent
\edef\ecrire{\write\toc{\string\par\string\itemitem
{\prefix\the\nosection.\the\nosubsection} {#1}
\string\leaderfill{\noexpand\number\pageno}}}\ecrire
}
\def\note #1{\advance\notenumber by 1
\footnote{$^{\the\notenumber}$}{\sevenrm #1}}

\def\nolabels{\def\wrlabel##1{}\def\eqlabel##1{}\def\reflabel##1{}}
\def\writelabels{\def\wrlabel##1{\leavevmode\vadjust{\rlap{\smash%
{\line{{\escapechar=` \hfill\rlap{\sevenrm\hskip.03in\string##1}}}}}}}%
\def\eqlabel##1{{\escapechar-1\rlap{\sevenrm\hskip.05in\string##1}}}%
\def\reflabel##1{\noexpand\llap{\noexpand\sevenrm\string\string\string##1}}}
\global\newcount\refno \global\refno=1
\newwrite\rfile
\def\ref{[\the\refno]\nref}
\def\nref#1{\xdef#1{[\the\refno]}\writedef{#1\leftbracket#1}%
\ifnum\refno=1\immediate\openout\rfile=\jobname.ref\fi
\global\advance\refno by1\chardef\wfile=\rfile\immediate
\write\rfile{\noexpand\item{#1\ }\reflabel{#1\hskip.31in}\pctsign}\findarg}
\def\findarg#1#{\begingroup\obeylines\newlinechar=`\^^M\pass@rg}
{\obeylines\gdef\pass@rg#1{\writ@line\relax #1^^M\hbox{}^^M}%
\gdef\writ@line#1^^M{\expandafter\toks0\expandafter{\striprel@x #1}%
\edef\next{\the\toks0}\ifx\next\em@rk\let\next=\endgroup\else\ifx\next\empty%
\else\immediate\write\wfile{\the\toks0}\fi\let\next=\writ@line\fi\next\relax}}
\def\striprel@x#1{}
\def\em@rk{\hbox{}}

\def\addref#1{\immediate\write\rfile{\noexpand\item{}#1}} 
\def\listrefs{
\ifnum\refno=1 \else
\immediate\closeout\rfile\writestoppt\baselineskip=14pt%
\vskip0pt plus.1\vsize\penalty-100\vskip0pt plus-.1
\vsize\bigskip\vskip\parskip\centerline{{\bf References}}\bigskip%
{\frenchspacing%
\parindent=20pt\escapechar=` \input \jobname.ref\vfill\eject}%
\nonfrenchspacing
\fi}
\def\startrefs#1{\immediate\openout\rfile=\jobname.ref\refno=#1}
\def\xref{\expandafter\xr@f}\def\xr@f[#1]{#1}
\def\refs#1{[\r@fs #1{\hbox{}}]}
\def\r@fs#1{\ifx\und@fined#1\message{reflabel \string#1 is undefined.}%
\xdef#1{(?.?)}\fi \edef\next{#1}\ifx\next\em@rk\def\next{}%
\else\ifx\next#1\xref#1\else#1\fi\let\next=\r@fs\fi\next}
%
\newwrite\lfile
{\escapechar-1\xdef\pctsign{\string\%}\xdef\leftbracket{\string\{}
\xdef\rightbracket{\string\}}}

\def\writestop{\def\writestoppt{\immediate\write\lfile{\string\pageno%
\the\pageno\string\startrefs\leftbracket\the\refno\rightbracket%
\string\def\string\secsym\leftbracket\secsym\rightbracket%
\string\secno\the\secno\string\meqno\the\meqno}\immediate\closeout\lfile}}
\def\writestoppt{}\def\writedef#1{}
\def\eqnn{\global\advance\neqno by 1 \ifinner\relax\else%
\eqno\fi(\eqprefix\the\neqno)}
%
\def\eqnd#1{\global\advance\neqno by 1 \ifinner\relax\else%
\eqno\fi(\eqprefix\the\neqno)\eqlabel#1
{\xdef#1{($\eqprefix\the\neqno$)}}
\edef\ewrite{\write\sym{\string\def\string#1{($\eqprefix%
\the\neqno$)}}%
}\ewrite%
}
%
\def\eqna#1{\wrlabel#1\global\advance\neqno by1
{\xdef #1##1{\hbox{$(\eqprefix\the\neqno##1)$}}}
\edef\ewrite{\write\sym{\string\def\string#1{($\eqprefix%
\the\neqno$)}}%
}\ewrite%
}
\def\em@rk{\hbox{}}
\def\xeqn{\expandafter\xe@n}\def\xe@n(#1){#1}
\def\xeqna#1{\expandafter\xe@na#1}\def\xe@na\hbox#1{\xe@nap #1}
\def\xe@nap$(#1)${\hbox{$#1$}}
\def\eqns#1{(\e@ns #1{\hbox{}})}
\def\e@ns#1{\ifx\und@fined#1\message{eqnlabel \string#1 is undefined.}%
\xdef#1{(?.?)}\fi \edef\next{#1}\ifx\next\em@rk\def\next{}%
\else\ifx\next#1\xeqn#1\else\def\n@xt{#1}\ifx\n@xt\next#1\else\xeqna#1\fi
\fi\let\next=\e@ns\fi\next}
\def\fig{fig.~\the\figno\nfig}
\def\nfig#1{\xdef#1{\the\figno}%
\immediate\write\sym{\string\def\string#1{\the\figno}}%
\global\advance\figno by1}%
\def\xfig{\expandafter\xf@g}\def\xf@g fig.\penalty\@M\ {}%
\def\figs#1{figs.~\f@gs #1{\hbox{}}}%
\def\f@gs#1{\edef\next{#1}\ifx\next\em@rk\def\next{}\else%
\ifx\next#1\xfig #1\else#1\fi\let\next=\f@gs\fi\next}%
\long\def\figure#1#2#3{\midinsert
#2\par
{\elevenpoint
\setbox1=\hbox{#3}
\ifdim\wd1=0pt\centerline{{\bf Figure\ #1}\hskip7.5mm}%
\else\setbox0=\hbox{{\bf Figure #1}\quad#3\hskip7mm}
\ifdim\wd0>\hsize{\narrower\noindent\unhbox0\par}\else\centerline{\box0}\fi
\fi}
\wrlabel#1\par
\endinsert}
\def\tab{table~\uppercase\expandafter{\romannumeral\the\tabno}\ntab}
\def\ntab#1{\xdef#1{\the\tabno}
\immediate\write\sym{\string\def\string#1{\the\tabno}}
\global\advance\tabno by1}
\long\def\table#1#2#3{\topinsert
#2\par
{\elevenpoint
\setbox1=\hbox{#3}
\ifdim\wd1=0pt\centerline{{\bf Table
\uppercase\expandafter{\romannumeral#1}}\hskip7.5mm}%
\else\setbox0=\hbox{{\bf Table
\uppercase\expandafter{\romannumeral#1}}\quad#3\hskip7mm}
\ifdim\wd0>\hsize{\narrower\noindent\unhbox0\par}\else\centerline{\box0}\fi
\fi}
\wrlabel#1\par
\endinsert}
\catcode`@=12
\def\draftend{\immediate\closeout\sym\immediate\closeout\toc
}
\draftstart
\preprint{T93/060}
\title{{\lq\lq Electroweak symmetry breaking through supersymmetry
breaking\rq\rq }}
\authorname{C.A. Savoy}
\address{\saclay}
\abstract
The connection between the scales of $ {\rm SU} (2)\times {\rm U}
(1) $ gauge symmetry
breaking and supersymmetry breaking is didactically displayed in the
framework of a T.O.Y. (Theory Overestimating Yukawas) model, a version of the
$ (M+1)SSM $ (supersymmetric extension of the standard model with a gauge
singlet) in which the relevant parameters are determined in the fixed point
regime. Some conspicuous features of supersymmetric particle physics are
reviewed in the light of this simplified model. An alternative theory
corresponding to lim $ (M+1)SSM \longrightarrow MSSM, $ leads to
interesting inequalities among
the supersymmetric breaking parameters of the $ MSSM. $
\endabstract
\vfill
{\it Invited talk at the 10th Capri Symposium on Particle Physics \lq\lq
Thirty
years of elementary particle theory\rq\rq}\par
{\it Capri, Italy}\par
{\it 1992}\par
\eject
\eject
\input mssymb

An instructive way to approach the issues in supersymmetric versions of the
standard model$ ^{[1,2]} $ is to analyse a simplified theory where the various
problems get disentangled enough to be dealt with one after the other. Now
let me briefly recall those issues. Supersymmetry has been mostly advocated
to solve the so-called hierarchy problem$ ^{[3]} $ so it is natural to assume
the theory not to be plagued with non-perturbative new physics up to the
scale of unification of the fundamental interactions, $ \Lambda_{ {\rm GUT}}.
$ The only new
scale should be that related to the spontaneous breaking of supersymmetry.
Alas we have so far no satisfactory mechanism to generate this breaking and
we have to rely on quite general arguments. Supersymmetry violations are then
parametrized by the so-called soft terms$ ^{[4]} $ in the effective theory$
^{[5]} $ below
unification. Within this framework for supersymmetric particle physics, the
main features I would like to emphasize are the following.

(i)\nobreak\ The parameters of the theory including the soft ones are likely
to be
related by the symmetries of the grand unified theory.

(ii)\nobreak\ The breaking of the electroweak gauge symmetry is presumably a
quantum
effect that is controlled by supersymmetry breaking and Yukawa couplings$
^{[6,7]}. $
The latter are also quite unknown but a relevant one should be the relatively
strong top-Higgs coupling.

(iii)\nobreak\ How is the Fermi scale of the $ {\rm SU(2)\times U(1)} $
breaking related to that of
supersymmetry breaking encoded in the soft terms?

(iv)\nobreak\ How massive can the (lightest) Higgs boson be in such theories
and can
we disprove them if the Higgs is not found at LEP2?

Guided by these questions I consider here an oversimplified version of the
theory discussed in Ref.$ [8]. $ This T.O.Y. (Theory Overestimating Yukawas)
model incorporates most of the features one requires from a realistic model,
e.g., the degrees of freedom. But it suffers twice my recourse to poetic
licence:

a)\nobreak\ gauge couplings are neglected in the RGE evolution of the
parameters that
are controlled by Yukawas;

b)\nobreak\ the unknown couplings are then assumed to be natural at the
unification
scale.

In order to make assumption b) more precise, we notice that in the
absence of gauge couplings the free parameters in the theory are controlled
by infrared fixed points, and are ultraviolet divergent.
The Yukawa couplings are then considered to be natural if they remain of the
same order of magnitude when they become large. The precise relations among
these parameters are consequences of the symmetries at the unification scale
that are not known. Remarkably enough, this theory possesses an attractive
fixed point in the evolution of ratios of parameters$ ^{[8]}. $  Whenever the
physics is controlled by these fixed ratio points, the unknown grand
unification relations among parameters are overwhelmed by new ones obtained
from the algebraic coefficients in the RGE. This non-trivial property leads
to the reduction of the unknown quantities to two fundamental ones: (the
order of magnitude of) the unification scale $ (\Lambda_{ {\rm GUT}}), $ which
is suggested by
the extrapolation of the gauge couplings, and supersymmetry breaking scale $
(\Lambda_{ {\rm SUSY}}) $
that will be eventually related to that of $ {\rm SU(2)\times U(1)} $ breaking
$ (G^{-1}_F). $ The
absence of gauge couplings in the RGE is not a fatal disease as far as they
can be turned on and the effects approximated in a simple way$ ^{[8]} $ but
the
T.O.Y. model has a serious drawback that I shall reveal... at the end.

The model must exhibit $ {\rm SU(3)\times SU(2)\times U(1)} $ gauge
supermultiplets (with gauge
boson and gaugino degrees of freedom) and chiral multiplets for the three
families of quarks-squarks and leptons-sleptons. I shall concentrate on the
heavy quarks of the third family that have the largest Yukawa couplings to
Higgs bosons: a SU(2) doublet of coloured chiral superfield $ (Q) $ including
the
left-handed quarks and their squarks and two coloured SU(2) singlet
superfields $ (T,B) $ holding the right-handed heavy quarks and their squarks.
Now it is well-known$ ^{[1]} $ that Higgs SU(2) doublets must be inserted in
pairs
of opposite hypercharge superfields, $ H_1 $ with $ Y=-1 $ and $ H_2 $ with $
Y=+1, $ to cancel
the higgsino contributions to the $ ABJ $ anomalies.

In supersymmetric theories$ ^{[9]}, $ the scalar and Yukawa interactions are
obviously related and are very tangibly embodied in a superpotential $ W, $
invariant under the gauge and global symmetries, such as $ B $ and $ L. $
Contrarily
to naive expectations, there is some room for introducing $ B $ or $ L $
violations
in supersymmetric theories$ ^{[10]}. $ This would lead to a rich interesting
phenomenology but I skip this issue here in.

Hence the supersymmetric Yukawa couplings of the Higgs to the heavy quarks
must correspond to the $ {\rm SU(3)\times SU(2)\times U(1)} $ invariant
superpotential:
$$ h_tTQH_2+h_bBQH_1 \eqno (1) $$
The two v.e.v.'s $ V_1= \left\langle H_1 \right\rangle $ and $ V_2=
\left\langle H_2 \right\rangle $ provide the scale of $ {\rm SU(2)\times U(1)}
$
breaking, $ V^2=V^2_1+V^2_2={1 \over 2 \sqrt{ 2}}G^{-1}_F, $ as well as the
quark masses: one known, $ m_b=h_bV_1, $
the other unknown, $ m_t=h_tV_2. $ I further assume here below $ h^2_t\gg
h^2_b, $ $ 1< \left(V_2/V_1 \right)= {\rm tan} \ \beta  \lesssim 6, $
to remain consistent with the fixed ratio point as discussed later. But the $
\beta $
-angle is actually unknown.

In the minimal supersymmetric version $ (MSSM) $ of the standard model$
^{[1,7,11-17]} $
the Higgs superfields must be coupled to each other in a supersymmetric mass
term of the form $ \mu H_1H_2. $ It is not surprising that the analysis of $
{\rm SU(2)\times U(1)} $
breaking$ ^{[11-17]} $ requires $ \mu \sim O(V). $ Notice, in particular, that
this coupling
(i.e., $ \mu \not= 0) $ is necessary$ ^{[7]} $ also to prevent a global U(1)
symmetry and its
spontaneous breaking. Now for the parameter $ \mu $ to be of $ O(V) $ it
should be
related to the supersymmetry breaking scale $ \Lambda_{ {\rm SUSY}} $ in spite
of the
supersymmetric nature of the $ \mu H_1H_2 $ coupling itself. It has been
suggested
that light Higgs are pseudo Goldstone bosons that survive the decoupling of
the heavy degrees of freedom after the breaking of global and local grand
unifying summetries$ ^{[18]}. $ It has also been pointed out that the $ \mu $
parameter
could arise as a relic interaction in the flat limit of broken supergravity$
^{[19]}. $
These ideas have some appeal but here we choose to replace the $ MSSM $ mass $
\mu $ by
the supersymmetric Yukawa couplings$ ^{[20-24]}. $
$$ \lambda SH_1H_2+{\kappa \over 3}S^3 \eqno (2) $$
where $ S $ is a gauge singlet supermultiplet. An unwanted U(1) global
symmetry
is avoided if $ \kappa \not= 0. $ These couplings together with (1) define a
superpotential
$ W \left(H_1,H_2,S,T,Q,... \right) $ trilinear in the chiral superfield.

Thus both $ W $ and the theory so defined (which I call $ (M+1)SSM $
hereafter) have
a manifest $ Z_3 $-symmetry which prevents hierarchy problems related to the
existence of light gauge singlets in supersymmetric theories$ ^{[25]}. $ Up to
now
the theory with the superpotential $ W $ given by (1) and (2) is
supersymmetric
and contains no explicit scale.

The spontaneous breaking of supergravity and its scale appear in the
effective theory$ ^{[5]} $ through the so-called soft terms, of $ O
\left(\Lambda_{ {\rm SUSY}} \right), $ which
nicely preserve the special renormalization pattern of
supersymmetric theories$ ^{[4]}. $ The appropriate ones in our case are the
following:

(a)\nobreak\ Analytic cubic interactions in the scalar potential
$$ A_\lambda \lambda SH_1H_2+A_\kappa{ \kappa \over 3}S^3+A_th_tH_iQT+...+
{\rm h.c.} , \eqno (3) $$
where the $ A $-parameters are of $ O \left(\Lambda_{ {\rm SUSY}} \right). $

(b)\nobreak\ Scalar mass terms
$$ m^2_S\mid S\mid^ 2+m^2_1 \left\vert H_1 \right\vert^ 2+m^2_2 \left\vert H_2
\right\vert^ 2+m^2_T\vert T\vert^ 2+m^2_Q\vert Q\vert^ 2+..., \eqno (4) $$
with $ m^2_i\sim O \left(\Lambda_{ {\rm SUSY}} \right) $ $
(i=S_1,H_1,H_2,...). $

(c)\nobreak\ Mass terms for the $ {\rm SU(3)\times SU(2)\times U(1)} $
gauginos, $ M_\alpha \left(\lambda^ \alpha \lambda^ \alpha \right), $ with $
M_\alpha \sim O \left(\Lambda_{ {\rm SUSY}} \right). $

So far the theory has many parameters only restricted by the low-energy
symmetries, $ {\rm SU(3)\times SU(2)\times U(1} ), $ $ B, L, $ etc. The
missing link in locally
supersymmetric theories is indeed the supersymmetry breaking mechanism. At
this stage, one can call into play the universality (or flavour independence)
of the supergravity couplings and assume a similar property for the soft
interactions (3)-(4) in the effective lagrangian with broken local
supersymmetry. However, at the quantum level, the scale dependence of the
lagrangian given by (1)-(4) has to be taken into account at least through the
(flavour dependent) RGE evolution of its couplings. Therefore the
universality conditions on the parameters of the soft interactions will be
broken by their scale dependence. Hence I assume flavour
independence as boundary conditions on the meaning parameters to be matched
at the unification scale $ \Lambda_{ {\rm GUT}}, $ not far from the
supergravity limit:
$$ A_a \left(\Lambda_{ {\rm GUT}} \right)=A_0,\ \ \ \ (a=t,\lambda ,x,...)
\eqno (5) $$
$$ \matrix{ m^2_i \left(\Lambda_{ {\rm GUT}} \right)  & =m^2_0,\ \ \
i=S_1H_1,H_2,T,Q,.... \cr M_\alpha \left(\Lambda_{ {\rm GUT}} \right)  &
=M_0,\ \ \ \alpha \in {\rm SU(3)\times SU(2)\times U(1)} \cr} \eqno  $$
Flavour conservation in neutral current interactions (FCNC) supplies some
phenomenological support for this universality, at least to some
approximation, in the quark-squark sector$ ^{[26]}. $

Though the precise choice of $ \Lambda_{ {\rm GUT}} $ would have some
quantitative impact on the
results here below, I disregard the fact that a few orders of magnitude
separate the unification of the gauge couplings from the Planck mass. Then
the parameters at the scale $ (V) $ of the breaking of the electroweak scale
are
obtained by solving the RGE for the gauge, Yukawa and soft coupling of the
theory$ ^{[27,16,21]}. $ On dimensional grounds the resulting soft terms must
take
the general forms:
$$ \matrix{ M_\alpha (V)  & ={g^2_\alpha (V) \over g^2_{\alpha_ 0}}M_0 \hfill
\cr A_a(V)  & =a_aA_0+c_aM_0 \hfill \cr m^2_i(V)  & =p_im^2_0+q_iA^2_0+\gamma_
iM^2_0+\delta_ i \left(A_0M_0 \right) \hfill \cr} \eqno (6) $$

The coefficients are functions of the gauge and Yukawa couplings. Since
several interactions of each type are mixed up in the RGE, simple solutions
might only arise in a fixed point regime. It has been shown in $ [8] $ that
when
gauge couplings are turned off (T.O.Y.) there is one attractive fixed ratio
point, i.e., a fixed point in the RGE of ratios of Yukawa couplings. This is
a non-trivial property of this particular T.O.Y. model. Its RGE for the
Yukawa couplings are as follows$ ^{[21]}: $
$$ \left( \matrix{\dot \lambda^ 2/\lambda^ 2 \cr\dot \kappa^ 2/\kappa^ 2
\cr\dot h^2_t/h^2_t \cr} \right)= \left( \matrix{ 4  & \ 1^{ }\   & 4 \cr 6  &
\ 3^{ }\   & 0 \cr 1  & \ 0^{ }_{ }\   & 8 \cr} \right) \left( \matrix{
\lambda^ 2 \cr 2\kappa^ 2_{ } \cr{ 3 \over 4}h^2_t \cr} \right) \eqno (7) $$
where the dot is for the derivative with respect to the scaling variable $ t=
\left( {\rm ln} \ \Lambda /8\pi^ 2 \right) $
and gauge coupling have been neglected as well as the Yukawa couplings of the
Higgs supermultiplet to those associated to the $ b $-quark and other lighter
fermions. The ratios $ \lambda^ 2:\kappa^ 2:h^2_t=1:{1 \over 2}:{4 \over 3} $
define an attractive fixed point in the
scale dependence of the ratios $ h^2_t/\lambda^ 2 $ and $ \kappa^ 2/\lambda^ 2
$ as given by the one-loop RGE
in (7). In order to check that these fixed ratio solutions are not a general
property of the running of Yukawa couplings in supersymmetric theories it is
enough to enlarge (7) to include the $ b $-quark Higgs coupling, $ h_b, $ and
verify
that there are no fixed ratio solutions anymore unless $ h_b=0. $

This critical point behaviour is equivalent to the requirement that Yukawa
couplings remain of the same order of magnitude in the ultraviolet region.
Large Yukawa couplings are relevant here as experiments require both the $ t $
-quark and the Higgs bosons to be relatively heavy.

Indeed, the obvious question is how heavy the Higgs bosons can be. Well, if
one considers the neutral scalar mass matrix for this theory and assumes the
right pattern of $ {\rm SU(2)\times U(1)} $ breaking, the mass of the lightest
Higgs is
bounded as follows$ ^{[22]}: $
$$ m^2_h\leq M^2_Z \left( {\rm cos}^22\beta +{\lambda^ 2 \over\bar g^2} {\rm
sin}^22\beta \right), \eqno (8) $$
where $ \bar g\simeq .53 $ is the gauge coupling of the $ Z $ boson. Although
the $ MSSM $ mentioned
here in before has a quite different vacuum structure, the corresponding
upper bound$ ^{[16]} $ on $ m_h $ is as in (8) with $ \lambda =0. $ This
tree-level bound would be
seriously affected by the radiative corrections in the presence of relatively
large values of $ \Lambda_{ {\rm SUSY}} $ but, still, larger $ \lambda $'s
could cause the lightest Higgs to be
heavier.

Now, Yukawa couplings grow logarithmically with the scale unless
small enough to be slowed down by the gauge couplings. In a sort it is
contradictory to allow for Yukawa getting out of the perturbative domain in a
supersymmetric theory. On the contrary, one should rather adopt the approach
of Ref.$ [28] $ and require all couplings to remain within their perturbative
region up to the unification scale, $ \Lambda_{ {\rm GUT}}. $ In this context
an upper bound can
be put on the $ \lambda $ coupling by taking all the others to be much
smaller, which
translates into a tree-level upper bound on the mass of the lightest Higgs,
to be corrected for quantum effects$ ^{[8,29]}. $ It can be further improved
by
taking into account the experimental lower bound on the top-Higgs coupling$
^{[30]}. $

In view of the relevance of this issue to future experiments, let me make a
parenthetical remark on radiative corrections to the bound, (8), on the
lightest Higgs mass$ ^{[31-35]}. $ The main contributions come from the fields
that
(i) are more coupled to the Higgs scalars, and (ii) present a large mass
splitting between the supersymmetric partners. In other words, the top-stop
gives the dominant terms. A rough even though often enough approximation to
the radiative shift of the lightest Higgs mass is obtained by stopping the RG
running of Higgs mass at an average stop mass and then using the SM running
to calculate the top quark effect on the Higgs mass$ ^{[32]}. $ This
approximation
has been also applied to the $ (M+1)SSM $ by several authors$
^{[36,37,8,29,30]}. $

An improved estimate of the radiative corrections consists in the evaluation
of the one-loop effective potential$ ^{[39]} $ and has been performed in the $
MSSM^{[33]} $
as well as in the framework of the $ (M+1)SSM^{[39]}. $ Complete one-loop
calculations have also been done$ ^{[35]} $ to check the previous
approximations for
the $ MSSM. $

All these detailed studies confirm that the radiative shift of the lightest
Higgs mass is a conspicuous effect in the presence of large supersymmetry
breaking effects. Thus the tree-level upper bound is increased to $ 140\ {\rm
GeV} $
according to those estimates. However, this fact should not discourage the
experimental search for the supersymmetric Higgs at LEP2. The scalar mass
matrix depends on various parameters of the supersymmetric models which are
constrained by two physical requirements:

(i)\nobreak\ the right pattern of $ {\rm SU(2)\times U(1)} $ breaking;

(ii)\nobreak\ the present experimental bounds on the supersymmetric particles.
The
spectrum of the $ (M+1)SSM $ has been analysed within these constraints$
^{[37]} $ and
the lightest Higgs scalar tend to be rather lighter than the upper bound. In
other words, LEP2 is really  going to exploit the bulk of the expected mass
range for (at least) the lightest Higgs.

After this long digression, let me turn back to the specific T.O.Y. model and
its \lq\lq natural\rq\rq\ simplifications. So I henceforth assume the fixed
ratio regime
of the T.O.Y. model. This fixes the Yukawa couplings at scales $ \Lambda \ll
\Lambda_ 0, $ $ \Lambda_ 0 $ being
defined, e.g., as the scale such that $ \lambda^ 2 \left(\Lambda_ 0
\right)\sim O(\pi ): $
$$ \matrix{ h^2_t(\Lambda )  & ={4 \over 3}\lambda^ 2(\Lambda ), \hfill \cr
\chi^ 2(\Lambda )  & ={1 \over 2}\lambda^ 2(\Lambda ), \hfill \cr \lambda^ 2
& = \left({9 {\rm ln} \left(\Lambda /\Lambda_ 0 \right) \over 8\pi^ 2}+{1
\over \lambda^ 2 \left(\Lambda_ 0 \right)} \right)^{-1} \hfill \cr} \eqno (9)
$$

Notice that the theory remain perturbative up to the unification scale if $
\Lambda_ 0>\Lambda_{ {\rm GUT}}. $
Taking $ \Lambda_{ {\rm GUT}}\sim 10^{16} {\rm \ GeV} $ one finds at a scale $
\Lambda \sim V=174\ {\rm GeV} , $ $ \lambda^ 2(V) \lesssim \bar g^2(V) $ and
the
tree-level bound $ m^2_h \lesssim M^2_Z. $ The corresponding value for $ h_t $
is meaningless because
this parameter is sensitive to strong gauge interactions that cannot be quite
neglected at low energies. An approximation has been introduced in Ref.$ [8] $
which shift its value to $ h_t(V)\simeq 1 $ and gives $ m_t\simeq 180( {\rm
sin} \ \beta ) {\rm GeV} . $ It is nothing to
be surprised about since by choosing values $ h^2_t \left(\Lambda_{ {\rm GUT}}
\right)>.1 $ at the unification
scale one always ends with $ m_t\gtrsim120( {\rm sin} \ \beta ) {\rm GeV} $ in
supersymmetric theories. (That
explains why $ m_t $ is a favourite prediction in recent papers).

Let us now turn our attention to one of the most important issues in
supersymmetric particle physics with the help of this simplified model. In
order to connect the electroweak scale with  $ \Lambda_{ {\rm SUSY}}, $ the
soft parameters at
low energies $ (\Lambda \sim V) $ must be expressed like in (6), in terms of $
A_0 $ and $ m^2_0 $
defined at $ \Lambda_{ {\rm GUT}} $ (in default of gauge couplings in the RGE
the coefficients of
$ M_0 $ vanish). First consider the analytic scalar couplings. The
correspondig
RGE in the T.O.Y. limit (gauge coupling neglected, fixed ratio regime for
Yukawas) turn out to be$ ^{[21]}: $
$$ \left( \matrix{\dot A_\lambda \cr\dot A_\kappa \cr\dot A_t \cr}
\right)={\dot \lambda^ 2 \over q} \left( \matrix{ 4  & \ 1_{ }\   & 4 \cr 6  &
\ 3_{ }\   & 0 \cr 1  & \ 0_{ }\   & 8 \cr} \right) \left( \matrix{ A_\lambda
\cr A_\kappa \cr A_t \cr} \right) \eqno (10) $$
Interestingly enough the boundary condition $ A_a \left(\Lambda_ 0 \right)=A_0
$ corresponds to an
eigenvector of the matrix above leading to the simple solution:
$$ \matrix{ A_\lambda (\Lambda )  & =A_\kappa (\Lambda )=A_t(\Lambda
)=\varepsilon (\Lambda )A_0 \hfill \cr \varepsilon (\Lambda )  & ={\lambda^
2(\Lambda ) \over \lambda^ 2 \left(\Lambda_ 0 \right)}={1 \over 1+{9\lambda^ 2
\left(\Lambda_ 0 \right) \over 8\pi^ 2} {\rm ln} \left({\Lambda_ 0 \over
\Lambda} \right)} \hfill \cr} \eqno (11) $$

The scalar mass parameters can be arranged in the convenient expressions:
$$ \matrix{ m^2_\lambda  &  =m^2_S+m^2_1+m^2_2-A^2_\lambda \hfill \cr
m^2_\kappa  &  =3m^2_S-A^2_\kappa \hfill \cr m^2_h  & =m^2_2+m^2_T+m^2_Q-A^2_t
\hfill \cr} \eqno (12) $$

In the T.O.Y. approximation the running of the mass parameters is then given
by$ ^{[21]}: $
$$ \matrix{ \left( \matrix{\dot m^2_\lambda \cr\dot m^2_\kappa \cr\dot m^2_h
\cr} \right)  & =\dot \lambda^ 2 \left( \matrix{ 4  & \ 1^{ }_{ }\   & 4 \cr 6
 & \ 3^{ }_{ }\   & 0 \cr 1  & \ 0^{ }_{ }\   & 8 \cr} \right) \left( \matrix{
m^2_\lambda \cr m^2_\kappa \cr m^2_h \cr} \right) \hfill \cr\dot m^2_2  &
=\dot m^2_1+3\dot m^2_Q \hfill \cr\dot m^2_T  & =2\dot m^2_Q \hfill \cr} \eqno
(13) $$
Again the boundary conditions $ (m^2_\lambda =m^2_k=m^2_h=3m^2_0-A^2_0) $
define a eigenvector of the
matrix in the RGE. Hence the mass parameters at low energies are given by
$$ \matrix{ m^2_1 \hfill&  =(2/3+\varepsilon /3)m^2_0 \hfill \cr m^2_2 \hfill&
 =(-2/3+5\varepsilon /3)m^2_0 \hfill \cr m^2_S \hfill&  =\varepsilon m^2_0
\hfill \cr m^2_T \hfill&  =(1/9+8\varepsilon /9)m^2_0 \hfill \cr m^2_Q \hfill&
 =(5/9+4\varepsilon /9)m^2_0 \hfill \cr} \eqno (14) $$
with $ \varepsilon $ given in (11).

As announced, the T.O.Y. hypotheses have led to a simple one-loop pattern
whose main characteristics are:

(i)\nobreak\ At low energies all the parameters are written in terms of the
high scale
$ \Lambda_ 0, $ where $ \lambda^ 2 \left(\Lambda_ 0 \right)\gg 1, $ and of $
m^2_0 $ and $ A_0 $ that measure $ \Lambda_{ {\rm SUSY}}. $ Consistency with
my
assumptions requires $ \Lambda_ 0\simeq \Lambda_{ {\rm GUT}}. $

(ii)\nobreak\ The $ A $-parameters, assumed to be all equal at $ \Lambda_{
{\rm GUT}} $ remain equal and
decrease as $ \left( {\rm ln} \ \Lambda /\Lambda^ 0 \right)^{-1} $ at lower
energies.

(iii)\nobreak\ The $ Y=-1 $ Higgs field gets a negative $ {\rm (mass)}^2, $ $
m^2_1 \longrightarrow - \left(2m^2_0/3 \right), $ which draws
$ {\rm SU(2)\times U(1)} $ breaking.

The soft terms are instrumental in breaking the electroweak gauge symmetry.
Models with $ {\rm SU(2)\times U(1)} $ breaking induced by the supersymmetric
part of the
scalar potential suffer from hierarchy problems$ ^{[25]}. $ On the contrary,
soft
cubic interactions and negative scalar mass terms are safe of these problems.
The real task is to check whether the resulting symmetry breaking pattern is
phenomenologically correct. Some of the questions to be investigated are the
following:

(a)\nobreak\ Coloured and/or charged scalar quark and leptons must have zero
v.e.v.'s,
an obstacle that I shall eventually meet with and discuss here below.

(b)\nobreak\ In models with more than one Higgs doublet, their v.e.v.'s have
to align
in such a way to preserve the electromagnetic gauge invariance. This is a
non-trivial property of supersymmetric theories with only Higgs doublets$
^{[40]}. $
The problem remains instead an open one in the presence of singlet fields
although a sufficient condition to prevent $ {\rm U(1)}_{ {\rm e.m.}} ${\bf\
}breaking is $ \lambda^ 2<g^2_2, $
where $ g_2 $ is the $ W $-boson coupling constant$ ^{[41]}. $

(c)\nobreak\ It is well-known that models with a richer Higgs sector have been
considered to produce spontaneous CP violation$ ^{[42]}. $ An elegant proof
has been
given$ ^{[43]} $ that softly broken supersymmetry models conserve CP. The
proof in
Ref.$ [43] $ assumes that $ {\rm U(1)}_{ {\rm e.m.}} $ is not broken but it
can be easily
generalized. Thus the v.e.v.'s can be taken to be real.

I postpone the discussion of (a) and skip the check that charged Higgs
v.e.v.'s are
zero in the T.O.Y. model with the parameters as calculated. The scalar
potential for the real (CP-even) parts of the neutral scalar field reads:
$$ \matrix{ V={\lambda^ 2 \over 2} \left[Y^4+2Y^2V^2_1+2Y^2V^2_2+2V^2_1V^2_2-2
\sqrt{ 2}Y^2V_1V_2 \right] \hfill \cr +{\bar g^2 \over 4} \left(V^2_1-V^2_2
\right)^2-A_0\varepsilon \lambda \left(2YV_1V_2+{ \sqrt{ 2} \over 3}Y^3
\right) \hfill \cr +\varepsilon m^2_0 \left(Y^2+V^2_1+V^2_2 \right)+{2 \over
3}(1-\varepsilon )m^2_0 \left(V^2_1-V^2_2 \right) \hfill \cr} \eqno (15) $$
where $ V_i= \left\langle {\rm Re} \ H^0_i \right\rangle $ $ (i=1,2) $ and $
Y=\langle {\rm Re} \ S\rangle . $ First consider the case $ \varepsilon \ll 1.
$

Then, the minization of the potential gives:
$$ \matrix{ V_1  & =Y=0 \hfill \cr V^2_2  & ={4m^2_0 \over 3\bar g^2},\ \ \ \
m^2_0={3 \over 4}M^2_Z \hfill \cr} \eqno (16) $$
Namely, a very simple relation between the supersymmetry breaking parameter $
m^2_0 $
and $ M^2_Z $ results under the T.O.Y. hypotheses. Since $ \lambda^
2\simeq\bar g^2, $ all neutral scalars
get a mass $ \simeq M_Z $ $ (M_W $ for the charged scalar).

However these results are inconsistent with phenomenology because leptons and
$ Q=-1/3 $ quarks remain massless as far as $ V_1=0. $ Some higgisinos remain
massless
as well. There is no way to improve this situation by taking into account
gauge interactions without invalidating our approximations (e.g. by assuming
very large gaugino masses). Before I discard this solution it is worth
stressing that, in spite of the relatively involved RGE and the presence of
many Yukawas, the existence of a fixed point reduces the relation between $
{\rm SU(2)\times U(1)} $
and supersymmetry breaking scales to the specific one in (16).

Since $ \varepsilon $ is not so small in practice, one may seek solutions such
that $ A_0\sim O(V/\varepsilon ), $
implying $ \Lambda_{ {\rm SUSY}}\gg V. $ The minimization is now a bit more
complicated but there
are indeed solutions with all neutral scalars taking a v.e.v. for particular
choices of $ A_0 $ and $ m^2_0. $ It is convenient to express these solutions
by the
relations:
$$ \matrix{ m^2_0={3 \over 4} \left(1+2\rho^ 2 \right)(- {\rm cos} \ 2\beta)
M^2_Z \hfill \cr \varepsilon A_0=2\rho \left({1 \over \sqrt{ 2}}+{1 \over
\sqrt{ 2}\rho^ 2+ {\rm sin} \ 2\beta} \right)M_Z \hfill \cr \rho^ 4(2 {\rm
sin} \ 2\beta -1)-2 \left( {\rm cos}^22\beta \right)\rho^ 2+ {\rm sin}^22\beta
=0 \hfill \cr \rho ={Y \over V},\ \ \ \ \ {\rm tan} \ \beta ={V_2 \over V_1},
\hfill \cr V= \sqrt{ V^2_1+V^2_2}\ \ \ \ \ M_Z=\bar gV\simeq \lambda V \hfill
\cr} \eqno (17) $$
Numerically they are as follows:

(a)\nobreak\ solutions interpolating between $ \left\{ {\rm tan} \ \beta =1,
\varepsilon A_0\simeq 2 \sqrt{ 2}M_Z, m^2_0\simeq 0 \right\} $ and $ \left\{
{\rm tan} \ \beta =10, \varepsilon A_0\simeq \sqrt{ 2}M_Z, m^2_0\simeq M^2_Z/2
\right\} ; $

(b)\nobreak\ solutions interpolating between $ \left\{ {\rm tan} \ \beta =-1,
\varepsilon A_0\simeq -1.5M_Z, m^2_0\simeq 0 \right\} $ and $ \left\{ {\rm
tan} \ \beta =-10, \varepsilon A_0\simeq -1.3M_Z, m^2_0\simeq M^2_Z/2 \right\}
; $

(c)\nobreak\ $ {\rm tan} \ \beta \simeq 1, \varepsilon A _0\simeq 4M_Z $ and $
m^2_0 $ between 0 and $ M^2_Z. $

In other words, there are solutions with large supersymmetry breaking, $
M_Z/\varepsilon <A_0<2M_Z/\varepsilon $
but they are characterized by relatively small values of $ m^2_0\sim O
\left(M^2_Z \right). $ By
including gauge couplings in the RGE and gaugino masses $ M_0 $ the T.O.Y.
model
could be turned into a realistic one. (As shown in Ref.$ [21], $ gaugino
masses
can also be generated by the decoupling of heavy multiplets of grand unified
theories in the presence of a large $ A_0). $

However, it is well-known that models of this kind are phenomenologically
excluded$ ^{[44,21]}. $ Indeed, if the supermultiplets holding the light
fermions
are taken into account with the corresponding soft terms fulfilling the
universal boundary conditions (5), there is an (approximate) condition to be
satisfied by the parameters $ A_0 $ and $ m^2_0. $ Let me formulate it by
considering the
inclusion of the electron supersymmetrized degrees of freedom, in particular
the scalars $ E, $ with $ Y=1, $ $ T=0 $ and $ L, $ with $ Y=-1/2, $ $ T=1/2.
$ The Yukawa coupling $ h_e=m_e/v_2\ll g_2 $
is very small. It has been shown$ ^{[21]} $ that the scalars $ E,L $ and $ H_1
$ will develop
a v.e.v. at low energies unless
$$ A^2_e<3 \left(m^2_1+m^2_E+m^2_L \right) \eqno (18) $$
at scales of $ O \left(A_0/\lambda_ e \right). $ Notice that (18) has been
proved only if the
parameters $ m^2_1,m^2_E $ and $ m^2_L $ are not very different (contrarily to
what is stated
in Ref.$ [45], $ that produces counterexamples by violating this assumption
clearly written in Ref.$ [21]). $ Therefore, models that violate (18) present
an
absolute minimum that breaks $ {\rm U(1)}_{ {\rm e.m.}} $ and $ L $ and are to
be discarded.
In the framework of the T.O.Y. model these parameters at low energy turn out
to be $ A_e\simeq 8A_0/9, $ $ m^2_1+m^2_E+m^2_L\simeq 8m^2_0/3. $ Hence, the
class of solutions with large
supersymmetric breaking violate (18) since $ A^2_0> \left(m^2_0/\varepsilon^ 2
\right). $ The inclusion of
gaugino masses cannot implement the right symmetry breaking pattern.

Notice that under similar assumptions, (18) also excludes models with $ {\rm
SU(2)\times U(1)} $
symmetry breaking occurring already at the tree-level. This would require$
^{[\ \ \ ]} $
$ A^2_\lambda >3 \left(m^2_1+m^2_2+m^2_S \right) $ and universality of the
soft terms would prevent (18). Thus,
radiative symmetry breaking is generally called for in model building.

Summing up, the simplifications introduced by the assumptions of the T.O.Y.
model allow for simple one-loop expressions for the various parameters. Gauge
symmetry breaking is induced because the $ {\rm (mass)}^2 $ of one of the
Higgs scalars
become negative at low energies under the effect of large Yukawa couplings in
its evolution, in particular that of the top coupling. In the T.O.Y. model
approximation $ \Lambda_{ {\rm SUSY}} $ can be analytically related to $ V. $
There is a minimum of
the scalar potential with $ M^2_Z=4m^2_0/3. $ In order to induce v.e.v.'s for
both
Higgs field one has to adjust otherwise the parameters. In general, the
relations will be loosened after the relaxation of the hypotheses of the
T.O.Y. model. But the existence of a phenomenologically satisfactory vacuum
is a strong constraint on the parameters, including the scale of
supersymmetry breaking.

A thorough analysis of the $ (M+1)SSM $ is in
progress$ ^{[46]} $ which takes into account present experimental limits on
supersymmetric particles. The solutions tend to favour large v.e.v.'s for the
singlet $ S. $ One finds $ \langle S\rangle =Y\gtrsim8M_Z, $ and relatively
large values of the
Yukawa couplings $ \kappa $ and $ \lambda $ are suppressed. In a sense one is
naturally led to
consider the following limit of the $ (M+1)SSM: $
$$ (Y/V) \longrightarrow \infty \ \ \ \ \kappa ,\lambda \longrightarrow 0
\eqno (19) $$
$ \mu =\lambda Y, $ $ \nu =\kappa Y $ fixed.

It is easily checked$ ^{[24,46]} $ that in this limit the singlet fields
decouple,
and the $ MSSM $ is approached. The singlet fermion, with mass $ 2\nu , $ and
the
singlet scalars, with masses $ 3A_\kappa \nu $ and $ \left(2\nu^ 2-m^2_S
\right) $ have couplings of $ O(V/Y) $ to
the $ MSSM $ sector. The usual soft supersymmetry breaking terms in the $ MSSM
$ are
then related to those in this limit of the $ (M+1)SSM $ as follows:
$$ \mu =\lambda Y\ \ \ \ B= \left(A_\lambda -\nu \right) \eqno (20) $$
as can be also seen from two of the potential minimization conditions in the
limit of Eq.(19):
$$ \eqalignno{ \left(m^2_1+m^2_2+2\mu^ 2 \right) {\rm sin} 2\beta &  =
\left(A_\kappa -\nu \right)\mu &  \cr \left(M^2_Z-m^2_1-m^2_2-2\mu^ 2 \right)
{\rm cos} 2\beta &  =m^2_2-m^2_1 & (21) \cr} $$
The minimization of the potential with respect to the singlet field requires:
$$ \nu ={A_\kappa \over 4} \left(1+ \sqrt{ 1-{8m^2_S \over A^2_\kappa}}
\right) \eqno (22) $$
and implies the following condition on the soft term parameters:
$$ A^2_\kappa >8m^2_S \eqno (23) $$
It becomes specially interesting if the universality conditions
(5) are required at the level of the $ (M+1)SSM. $ Since $ \kappa $ and $
\lambda $ are small, I can
use the analytic approximate solutions of Ref.$ [17] $ to write the low energy
soft interactions in terms of the universal parameters $ A_0, $ $ m^2_0 $ and
$ M_0, $ as
follows:
$$ \matrix{ A_h \hfill&  =A_0 \left(1-.78h^2 \right)+M_0 \left(3.96-1.72h^2
\right) \hfill \cr A_\lambda \hfill&  =A_0 \left(1-.39h^2 \right)+M_0
\left(.59-.86h^2 \right) \hfill \cr A_\kappa \hfill&  \simeq A_0 \hfill \cr
m^2_{H_1} \hfill&  =m^2_0+.52M^2_0 \hfill \cr m^2_{H_2} \hfill&
=m^2_{H_1}-h^2 \left[1.16m^2_0+ \left(.384-.30h^2 \right)A^2_0 \right. \hfill
\cr  \hfill& \left.+ \left(1.72-1.33h^2 \right)M_0A_0+ \left(4.93-1.48h^2
\right)M^2_0 \right] \hfill \cr m^2_S \hfill&  =m^2_0 \hfill \cr} \eqno (24)
$$
where $ h^2 $ is the top-Higgs Yukawa coupling at low energies. (Analogous
expressions are obtained for the other soft terms). Then, the inequalities
(18) and (25) can be translated into the following inequalities$ ^{[46]} $ for
$ A_0,M_0,m^2_0: $
$$ \sqrt{ 8}m_0< \left\vert A_0 \right\vert <.69M_0+ \sqrt{ 3.25M^2_0+9m^2_0}
\eqno (25) $$
On the top of (25) one has to fulfill (21) to avoid a $ {\rm SU(2)\times U(1)}
$ conserving
solution. Two remarks are in order here:

(a)\nobreak\ This relation follows from the minimization with respect to the
singlet
field, which has no equivalent in the $ MSSM $

(b)\nobreak\ It is only an approximation as far as corrections $ O(V/Y) $ are
taken into
account. In general$ ^{[46]} $ the solutions of the $ (M+1)SSM $ can depart
quite a lot
from this asymptotic pattern.

It is worth noticing that even near the $ MSSM $ the $ (M+1)SSM $ has
additional
degrees of freedom that mix at $ O(V/Y) $ to the neutralino and neutral Higgs
sector and might be observable. In the limiting cases (21)-(24) yield $ \nu
\simeq A_0/4 $
so that a $ CP=+1 $ state could be relatively light.

Recent experimental data have provided useful lower bounds on the parameters
of supersymmetric models. Only lately the experiments reached an energy scale
that can be considered as natural in supersymmetric theories. In view of that
most of the theoretical analyses performed a few years ago must be revisited.
This has been done to a large extent in the framework of the $ MSSM $ and has
proved useful in the analysis of LEP data. A similar effort is now under way
in the case of the $ (M+1)SSM. $

At the First Capri Symposium, in 1983, I reviewed supersymmetric particle
physics$ ^{[2]} $ and concluded that in spite of the striking effects that are
expected the natural mass spectrum seemed to be such that no significant
experimental result should be expected before LEP. (I underestimated slightly
the ability of hadron colliders to measure gluino masses). LEP1 is already an
old story. But the great excitement is still to come with the next generation
of experiments, (LEP2, LHC), supposed to test the mostly natural region of
masses of supersymmetric particles.
\vfill\eject
\centerline{{\bf REFERENCES}}
\vskip 20mm
\item{$\lbrack$1$\rbrack$}For a review see, for example, H. Nilles, Phys. Rep.
C {\bf 110} (1984)
1.

\item{$\lbrack$2$\rbrack$}C.A. Savoy, Lectures at the 1st Capri Symposium and
at the 18th
Rencontre de Moriond, in {\sl Beyond the Standard Model;\/} J.T.T. Van, ed.,
(Editions Fronti\`eres, Gif, 1983).

\item{$\lbrack$3$\rbrack$}L. Maiani in Proc. Gif Summer School, 1979 (IN2P3,
Paris, 1980).
\item{\nobreak\ \nobreak\ \nobreak\ }M. Veltman, Acta Phys. Pol. B {\bf 12}
(1981) 437.
\item{\nobreak\ \nobreak\ \nobreak\ }E. Witten, Nucl. Phys. B {\bf 188} (1981)
513.

\item{$\lbrack$4$\rbrack$}L. Girardello and M.T. Grisaru, Nucl. Phys. B {\bf
194} (1982) 65.

\item{$\lbrack$5$\rbrack$}R. Barbieri, S. Ferrara and C.A. Savoy, Phys. Lett.
{\bf 119B} (1982)
343.
\item{\nobreak\ \nobreak\ \nobreak\ }L. Hall, J. Likken and S. Weinberg, Phys.
Rev. D {\bf 27} (1983) 2359.

\item{$\lbrack$6$\rbrack$}L. Ibanez and G.G. Ross, Phys. Lett. {\bf 110B}
(1982) 215.

\item{$\lbrack$7$\rbrack$}L. Alvarez-Gaume, J. Polshinsky and M. Wise, Nucl.
Phys. B {\bf 221}
(1983) 495.

\item{$\lbrack$8$\rbrack$}P. Bin\'etruy and C.A. Savoy, Phys. Lett. B {\bf
277} (1992) 453.

\item{$\lbrack$9$\rbrack$}For a review see, e.g., P. Fayet and S. Ferrara,
Phys. Rep. {\bf 32C}
(1977) 250.

\item{$\lbrack$10$\rbrack$}S. Dimopoulos and L. Hall, Phys. Lett. B {\bf 207}
(1988) 210 and
references therein.

\item{$\lbrack$11$\rbrack$}L.E. Ibanez and C. Lopez, Nucl. Phys. B {\bf 233}
(1984) 511.

\item{$\lbrack$12$\rbrack$}J. Ellis, J.S. Hagelin, D.V. Nanopoulos and K.
Tamvakis, Phys.
Lett. {\bf 125B} (1983) 275.
\item{\nobreak\ \nobreak\ \nobreak\ \nobreak\ }E. Cremmer, P. Fayet, L.
Girardello, Phys. Lett. {\bf 122B} (1983) 346.

\item{$\lbrack$13$\rbrack$}M. Claudson, L. Hall and I. Hinchliffe, Nucl. Phys.
B {\bf 228} (1983)
501.

\item{$\lbrack$14$\rbrack$}C. Kounnas, A.B. Lahanas, D.V. Nanopoulos and M.
Quiros, Nucl.
Phys. B {\bf 236} (1984) 438.

\item{$\lbrack$15$\rbrack$}A. Bouquet, J. Kaplan and C.A. Savoy, Phys. Lett.
{\bf 148B} (1984) 69.

\item{$\lbrack$16$\rbrack$}K. Inoue, A. Kakuto, H. Komatsu and S. Takeshita,
Prog. Theo.
Phys. {\bf 67} (1982) 1859; {\bf 68} (1982) 927.

\item{$\lbrack$17$\rbrack$}A. Bouquet, J. Kaplan and C.A. Savoy, Nucl. Phys. B
{\bf 262} (1985)
299.

\item{$\lbrack$18$\rbrack$}K. Inoue, A. Kakuto and T. Takano, Prog. Theo.
Phys. {\bf 75} (1986)
664.
\item{\nobreak\ \nobreak\ \nobreak\ \nobreak\ }A.A. Ansel'm and A.A. Johansen,
Phys. Lett. B {\bf 200} (1988) 331.

\item{$\lbrack$19$\rbrack$}G.F. Giudice and A. Masiero, Phys. Lett. B {\bf
206} (1988) 480.

\item{$\lbrack$20$\rbrack$}H.P. Nilles, M. Srednicki and D. Wyler, Phys. Lett.
{\bf 120B} (1983)
346.

\item{$\lbrack$21$\rbrack$}J.P. Derendinger and C.A. Savoy, Nucl. Phys. B
{\bf 237} (1984) 307.

\item{$\lbrack$22$\rbrack$}M. Drees, Intern. J. Mod. Phys. A {\bf 4} (1989)
3635.

\item{$\lbrack$23$\rbrack$}L. Durand and J. Lopez, Phys. Lett. B {\bf 217}
(1989) 463.

\item{$\lbrack$24$\rbrack$}J. Ellis, H. Haber, J.F. Gunion, L. Roszhowski and
F. Zwirner,
Phys. Rev. D {\bf 39} (1989) 844.

\item{$\lbrack$25$\rbrack$}H.P. Nilles, M. Srednicki and D. Wyler, Phys. Lett.
B {\bf 124} (1983)
337.
\item{\nobreak\ \nobreak\ \nobreak\ \nobreak\ }A. Lahanas, Phys. Lett. B {\bf
124} (1983) 341.

\item{$\lbrack$26$\rbrack$}S. Dimopoulos and H. Georgi, Nucl. Phys. B {\bf
193} (1981) 150.
\item{\nobreak\ \nobreak\ \nobreak\ \nobreak\ }J. Ellis and D.V. Nanopoulos,
Phys. Lett. B {\bf 110} (1982) 44.
\item{\nobreak\ \nobreak\ \nobreak\ \nobreak\ }R. Barbieri and R. Gatto, Phys.
Lett. B {\bf 110} (1982) 211.

\item{$\lbrack$27$\rbrack$}R. Barbieri, S. Ferrara, L. Maiani, F. Palumbo and
C.A. Savoy,
Phys. Lett. B {\bf 115} (1982) 212.
\item{\nobreak\ \nobreak\ \nobreak\ \nobreak\ }M. Einhorn and D.R.T. Jones,
Nucl. Phys. B {\bf 196} (1982) 475.

\item{$\lbrack$28$\rbrack$}N. Cabibbo, L. Maiani, G. Parisi and R. Petronzio,
Nucl. Phys. B
{\bf 158} (1979) 295.

\item{$\lbrack$29$\rbrack$}J. Espinoza and M. Quiros, Phys. Lett. B {\bf 279}
(1992) 92.

\item{$\lbrack$30$\rbrack$}J. Espinoza and M. Quiros, Instituto de Estructura
de la Materia
preprint IEM-FT-60/92.
\item{\nobreak\ \nobreak\ \nobreak\ \nobreak\ }U. Ellwanger and M. Lindner,
Heidelberg preprint
HD-THEP-92-48rev.
\item{\nobreak\ \nobreak\ \nobreak\ \nobreak\ }T. Elliot, S.F. King and P.L.
White, Southampton preprint SHEP 92/93-11.

\item{$\lbrack$31$\rbrack$}Y. Ohada, M. Yamaguchi and T. Yanagida, Prog. Theo.
Phys. Lett. {\bf 85}
(1991) 1.
\item{\nobreak\ \nobreak\ \nobreak\ \nobreak\ }H. Haber and R. Hempfling,
Phys. Rev. Lett. {\bf 66} (1991) 1815.
\item{\nobreak\ \nobreak\ \nobreak\ \nobreak\ }J. Ellis, G. Ridolfi and F.
Zwirner, Phys. Lett. B {\bf 257} (1991) 83.

\item{$\lbrack$32$\rbrack$}R. Barbieri, M. Frigeni and F. Caravaglios, Phys.
Lett. B {\bf 258}
(1991) 167.

\item{$\lbrack$33$\rbrack$}J. Ellis, R. Ridolfi and F. Zwirner, Phys. Lett. B
{\bf 262} (1991) 477.

\item{$\lbrack$34$\rbrack$}A. Yamada, Phys. Lett. B {\bf 263} (1991) 233.

\item{$\lbrack$35$\rbrack$}P. Chankowski, S. Pokorski and J. Rosiek, Phys.
Lett. B {\bf 274} (1992)
191.
\item{\nobreak\ \nobreak\ \nobreak\ \nobreak\ }A. Brignole, Phys. Lett. B
{\bf 281} (1992) 284.

\item{$\lbrack$36$\rbrack$}J. Espinoza and M. Quiros, Phys. Lett. B {\bf 266}
(1991) 389.

\item{$\lbrack$37$\rbrack$}U. Ellwanger and M. Rausch de Traubenberg, Z. Phys.
C {\bf 53} (1992)
521.

\item{$\lbrack$38$\rbrack$}S. Coleman and E. Weinberg, Phys. Rev. D {\bf 7}
(1973) 1888.
\item{\nobreak\ \nobreak\ \nobreak\ \nobreak\ }S. Weinberg, Phys. Rev. D {\bf
7} (1973) 2887.

\item{$\lbrack$39$\rbrack$}U. Ellwanger, Heidelberg preprint HD-THEP-93-4.

\item{$\lbrack$40$\rbrack$}D. Chang and A. Kumar, Phys. Rev. D {\bf 31} (1985)
2698.

\item{$\lbrack$41$\rbrack$}M. Olechowski and S. Pokorski, Phys. Lett. B {\bf
214} (1988) 393.

\item{$\lbrack$42$\rbrack$}G.C. Branco, Phys. Rev. D {\bf 22} (1980) 201.

\item{$\lbrack$43$\rbrack$}J.C. Romao, Phys. Lett. B {\bf 173} (1986) 309.

\item{$\lbrack$44$\rbrack$}J.M. Fr\`ere, D.R.T. Jones and S. Raby, Nucl. Phys.
B {\bf 222} (1983) 11.

\item{$\lbrack$45$\rbrack$}J.F. Gunion, H.E. Haber and M. Sher, Nucl. Phys. B
{\bf 306} (1988) 1.

\item{$\lbrack$46$\rbrack$}U. Ellwanger et al., in preparation

\listrefs
\draftend
\end